\documentclass[aps,prd,onecolumn,superscriptaddress]{revtex4}  % for review and submission 
\usepackage{graphicx}  % needed for figures
\usepackage{dcolumn}   % needed for some tables
\usepackage{bm}        % for math
\usepackage{amssymb}   % for math
\usepackage{amsmath}   % for math
\usepackage{multirow}
\usepackage{color}
\usepackage{units}

\newcommand{\etal}{{\it et al.}}

\newcommand{\numu}{\ensuremath{\nu_{\mu}}}                   % nu_mu
\newcommand{\nue}{\ensuremath{\nu_{e}}}                      % nu_e
\newcommand{\anue}{\ensuremath{\overline{\nu}_{e}}}          % anu_e
\newcommand{\anumu}{\ensuremath{\overline{\nu}_{\mu}}}       % anu_mu

\newcommand{\nova}{NO$\nu$A}
\newcommand{\ra}{\,\mathord{\rightarrow}\,}

\begin{document}

\title{Prospects for measurement of the neutrino mass hierarchy}
\author{R.~B.~Patterson}
% NOTE: affil line note rendering properly for me; commenting out for my own built PDF version to keep rest of document sensible
\affiliation{Division of Physics, Mathematics, and Astronomy, California Institute of Technology, Pasadena, California 91125; email: rbpatter@caltech.edu}

\begin{abstract}
The unknown neutrino mass hierarchy -- whether the $\nu_3$ mass eigenstate is the heaviest or the lightest -- represents a major gap in our knowledge of neutrino properties.  Determining the hierarchy is a critical step toward further precision measurements in the neutrino sector.  The hierarchy is also central to interpreting the next generation of neutrinoless double beta decay results, plays a role in numerous cosmological and astrophysical questions, and serves as a powerful model discriminant for theories of neutrino mass generation and unification.  Various current and planned experiments claim sensitivity for establishing the neutrino mass hierarchy.  We review the most promising of these here, paying special attention to points of concern and consolidating the projected sensitivities into an outlook for the years ahead.
\end{abstract}

\maketitle

\section{INTRODUCTION}
In the Standard Model with non-zero neutrino masses, the three neutrino flavor eigenstates $\nu_e$, $\nu_\mu$, and $\nu_\tau$ are non-trivial linear combinations of the three mass eigenstates $\nu_1$, $\nu_2$, and $\nu_3$.  The complex matrix $U$ relating the flavor and mass eigenstates to each other, called the PMNS matrix for the authors of Refs.~\cite{pontecorvo1,pontecorvo2,mns}, can be parametrized in terms of three rotation angles and three complex phases:
\begin{equation}\label{equ:pmns}
U=
\left(\begin{array}{ccc}
1&0&0\\
0& c_{23}&s_{23}\\
0&-s_{23}&c_{23}
\end{array}\right)
\left(\begin{array}{ccc}
 c_{13}&0&s_{13}e^{-i\delta}\\
0&1&0\\
-s_{13}e^{i\delta}&0&c_{13}
\end{array}\right)
\left(\begin{array}{ccc}
 c_{12}&s_{12}&0\\
-s_{12}&c_{12}&0\\
0&0&1
\end{array}\right)
\left(\begin{array}{ccc}
e^{i\alpha_1/2}&0&0\\
0&e^{i\alpha_2/2}&0\\
0&0&1
\end{array}\right)\ \ ,
\end{equation}
with $c_{ij}\mathord{=}\cos\theta_{ij}$ and $s_{ij}\mathord{=}\sin\theta_{ij}$.  While this neutrino mixing gives rise to several physical phenomena, the most experimentally fruitful one so far has been neutrino oscillations, whereby the flavor composition of a neutrino can vary as it propagates.  The probability $P$ that a relativistic neutrino produced with flavor $l$ will be detected as a neutrino of flavor $l'$ after traveling a distance $L$ through vacuum is given by
\begin{equation}\label{equ:oscprob}
\begin{split}
P(\nu_l\ra\nu_{l'})\ =\ \delta_{ll'}\;&-\;4\sum_{i>j}\Re(U^*_{l i}U_{l' i}U_{l j}U^*_{l' j})\sin^2\left(\frac{\Delta m^2_{ij}L}{4E}\right)\\
                                                   &+\;2\sum_{i>j}\Im(U^*_{l i}U_{l' i}U_{l j}U^*_{l' j})\sin\left(\frac{\Delta m^2_{ij}L}{2E}\right)\ ,
\end{split}
\end{equation}
where $E$ is the neutrino's energy, $\Delta m^2_{ij}\equiv m^2_{i}-m^2_{j}$ is the difference between the squares of the masses of the mass eigenstates $\nu_i$ and $\nu_j$, $\delta_{ll'}$ is the Kronecker delta, and the matrix elements $U_{li}$ are defined by
\begin{equation}
U=\left(\begin{array}{ccc}
U_{e1}&U_{e2}&U_{e3}\\
U_{\mu 1}&U_{\mu 2}&U_{\mu 3}\\
U_{\tau 1}&U_{\tau 2}&U_{\tau 3}
\end{array}\right)\ .
\end{equation}

Our quantitative understanding of the neutrino sector has grown tremendously in the past two decades primarily due to neutrino oscillation experiments.  The ``solar'' mass splitting $\Delta m^2_{21}$~\cite{gando} and the thirty-times-larger ``atmospheric'' mass splitting $\Delta m^2_{32}$~\cite{minos3nu} have been measured with impressive precision:
\begin{align}
\Delta m^2_{21}&=(7.53\pm0.18)\mathord{\times}10^{-5}\ \mathrm{eV}^2\label{equ:solarsplit}\\
\Delta m^2_{32}&=(2.34\pm0.09)\mathord{\times}10^{-3}\ \mathrm{eV}^2\ \ [\mathrm{NH}]\ \ \ \mathrm{or}\ \ \ (-2.37^{+0.07}_{-0.11})\mathord{\times}10^{-3}\ \mathrm{eV}^2\ \ [\mathrm{IH}]\ .\label{equ:atmsplit}
\end{align}
The normal hierarchy (NH) and inverted hierarchy (IH) distinction is defined below.\footnote{$\Delta m^2_{31}$ is trivially related to the other two splittings.}  In addition to these mass splittings, we know at varying levels of precision the values of the three mixing angles $\theta_{12}$, $\theta_{13}$, and $\theta_{23}$~\cite{pdg}.  We do {\em not} have good information about the sign of the atmospheric mass splitting, the complex phases in the PMNS matrix of Eq.~(\ref{equ:pmns}), the absolute masses $\{m_i\}$ of the neutrinos, or whether neutrinos are Majorana particles.  Furthermore, several experimental anomalies suggest that extensions to the neutrino standard model may be needed~\cite{steriles}.

\begin{figure}[hbtp]
\begin{center}
  \includegraphics[width=0.7\linewidth]{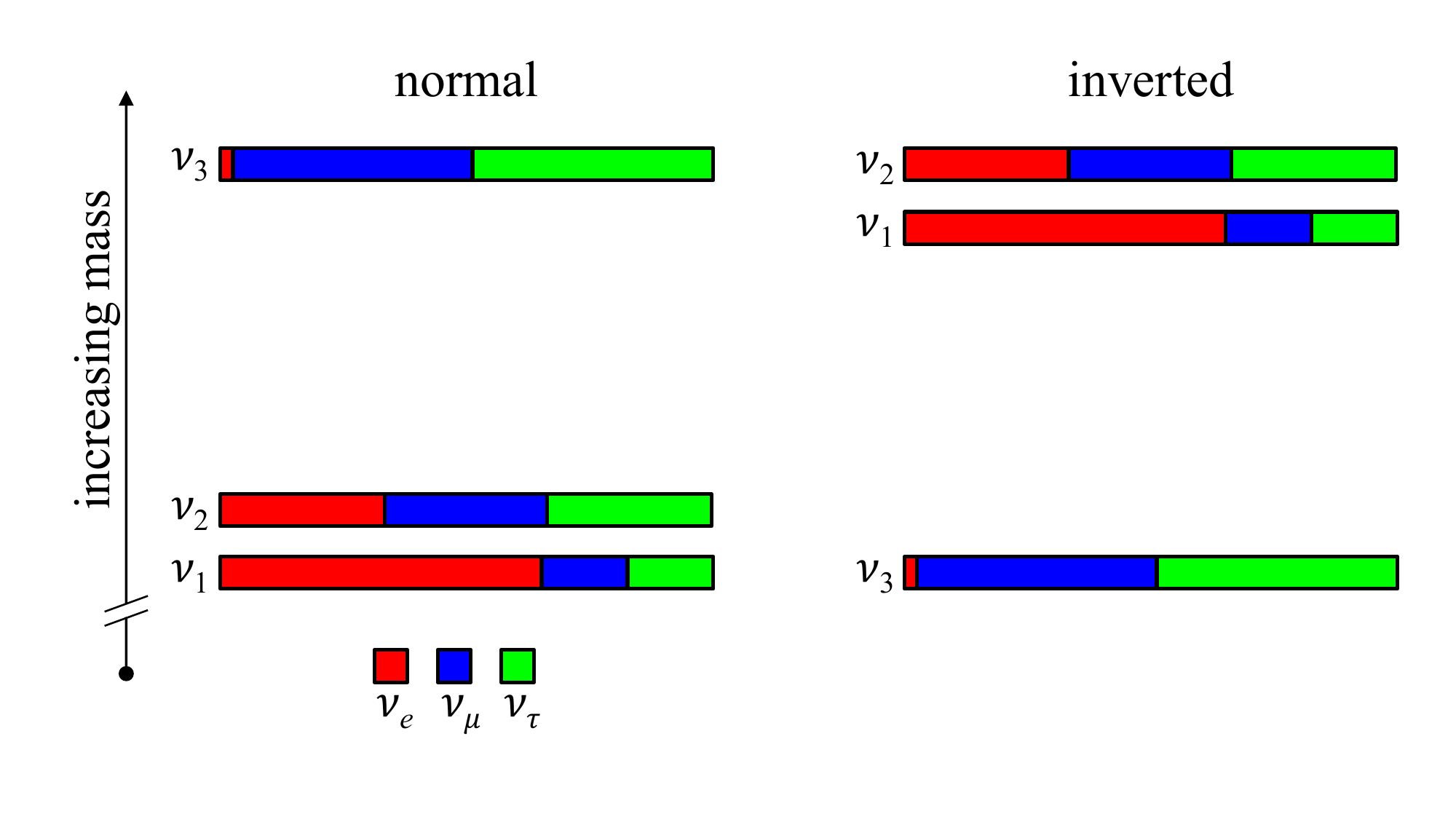}
  \caption{The two possible neutrino mass hierarchies.  The colors represent the approximate flavor admixtures present in each mass eigenstate.  In the normal case, the masses exhibit a hierarchical pattern, and the lightest neutrino has the largest admixture of electron flavor.  The relative ordering $m_2\mathord{>}m_1$ is known through observations of solar neutrinos, which are subject to resonant matter effects in the sun~\cite{gando,wolfenstein,ms1,ms2}.}
  \label{fig:hier}
\end{center}
\end{figure}

The scope of this review is limited to the experimental prospects for measuring the unknown sign of the atmospheric mass splitting or, equivalently, whether $\nu_3$ is the heaviest or lightest mass eigenstate.  These two possibilities are termed, respectively, the normal and inverted neutrino mass hierarchies, and Figure~\ref{fig:hier} shows the two cases graphically.  Some authors prefer the term ``mass ordering'' over ``mass hierarchy'' when a third case -- a degenerate spectrum -- is relevant to the context.  In a degenerate spectrum, the mass of the lightest neutrino is large compared to the individual mass differences, {\em i.e.}\ $m_1\,\mathord{\approx}\, m_2\,\mathord{\approx}\, m_3$.  Of note, neutrinoless double beta decay rates depend strongly on the mass hierarchy only if the mass spectrum is non-degenerate~\cite{0vbb}.  In more general contexts, authors regularly use ``ordering'' and ``hierarchy'' interchangeably.

The precision of neutrino sector measurements has reached a point where the unknown hierarchy is a major hurdle to further progress.  Knowledge of the hierarchy is also vital to interpretation of neutrinoless double beta decay results, serves as an input to cosmological and astrophysical measurements, and is a powerful discriminant among unification and neutrino mass models.  The neutrino mass hierarchy may also be the next major unknown of the Standard Model to be measured.  In this article we review the experimental outlook for making this measurement.

\section{MAKING THE HIERARCHY MEASUREMENT}

\subsection{Physical principles}
Several distinct approaches for probing the mass hierarchy are in use or under consideration by current and future experiments.  One approach is motivated by the vacuum oscillation probability in Eq.~(\ref{equ:oscprob}).  Since this probability depends on each of the mass splittings $\Delta m^2_{ij}$, a sufficiently precise measurement of $P(E)$ for a given set of initial and final flavors $l$ and $l'$ can determine whether $\Delta m^2_{32}$ is slightly larger or smaller than $\Delta m^2_{31}$.  This approach suits reactor-sourced $\overline{\nu}_e\ra\overline{\nu}_e$ oscillation measurements, as low energy neutrinos (few MeV) are needed to obtain significant contributions from the $\Delta m^2_{21}$ terms at distances $L$ that are short enough (tens of kilometers) to still have useful event rates at the detector.

For neutrinos traveling through matter, the presence of electrons in the propagation medium modifies the Hamiltonian for $\nu_e$ differently than it does for $\nu_\mu$ and $\nu_\tau$~\cite{wolfenstein,ms1,ms2,barger}.  Qualitatively, the effective masses of the $\nu_e$-dominated $\nu_1$ and $\nu_2$ states are increased (or decreased for antineutrinos) by the presence of the electrons, and the kinematic phases $\Delta m^2_{ij}L/(4E)$ follow suit.  This either compresses or expands the effective mass splittings, depending on the hierarchy, modifying the oscillation probabilities in a observable way.

In a different arena, cosmological measurements of large-scale structure, the cosmic microwave background, and its $B$-mode polarization provide access to the hierarchy via the sum of the neutrino masses~\cite{cosmo,cosmo2}.  In particular, were these data to yield $\sum m_i\lesssim 2\sqrt{|\Delta m^2_{32}|}\sim 100\ \mathrm{meV}$, the hierarchy would have to be normal.  A analogous constraint can be made for the effective electron neutrino mass measured in beta decay experiments, but upcoming experiments will not have the mass sensitivity required to provide hierarchy information even if the necessary conditions on the mass spectrum (normal hierarchy and small $m_1$) are met.

Neutrinoless double beta decay searches will rely on external knowledge of the hierarchy to interpret observed (or limited) decay rates~\cite{0vbb} and will not provide independent measurements of the hierarchy themselves.  Likewise, the neutrino fluxes from supernovae carry signatures of the hierarchy due to both matter effects and collective (high neutrino density) effects~\cite{supernova1,supernova2,supernova3}, but uncertainties in the astrophysical models make reliable determination of the hierarchy through supernova neutrino observations unlikely even if a nearby supernova event were to occur in the near future.

In the sections below, we describe specific experimental efforts toward measuring the mass hierarchy, discussing the advantages, possible pitfalls, and expected sensitivity of each.  The collection is not exhaustive in that we exclude experiments that do not have notable hierarchy sensitivity or are unlikely to proceed due to programmatic, budgetary, or other practical considerations.

\subsection{Statistical issues}
First, it is worth devoting a few words to the statistical interpretation of the sensitivities.  Though the question at hand can be stated quite simply -- Is the hierarchy normal or inverted? -- much discussion of its statistical treatment has taken place in recent years, largely due to the lack of a universally accepted definition of sensitivity.  Three definitions appear most often in the literature.

The first makes use of pseudo-data formed from the predictions for the experimental observables.  More specifically, one of the hierarchies is taken as truth and the corresponding predicted observables are used directly as the pseudo-data, with no statistical or systematic fluctuations introduced.  This data set, for which the name ``Asimov'' data set~\cite{asimov} is gaining traction in the literature, is then fit assuming the opposite hierarchy to find the minimum $\chi^2$ for this wrong hierarchy assumption, marginalizing over all free oscillation and systematic nuisance parameters.  Since the correct assumption must produce $\chi^2\mathord{=}0$ for this data set, the minimum $\chi^2$ found for the wrong hierarchy represents a typical $\Delta\chi^2$ between the hypotheses.  Motivated by the behavior of Gaussian systems, $\sqrt{\Delta\chi^2}$ is then taken as the hierarchy determination sensitivity as a number of $\sigma$.

In a second approach, an ensemble of fully fluctuated data sets is simulated assuming one hierarchy and used to calculate the expectation value for the $\Delta \chi^2$ between the incorrect and correct hierarchy hypotheses, again marginalizing over nuisance parameters for each simulated data set.  The sensitivity is then quoted as $\sqrt{\langle\Delta \chi^2\rangle}$.

The third common approach starts again with an ensemble of fluctuated data sets assuming one hierarchy.  The sensitivity is quoted as the frequency with which the correct hierarchy hypothesis fits to a lower $\chi^2$ than the incorrect one.  This sensitivity may be given as a $p$-value directly or converted into the corresponding number of Gaussian $\sigma$.

Many variants of these approaches have also been used or proposed.  A given author may choose one method over another to facilitate comparisons or combinations with past research, to promote a Bayesian or frequentist philosophy to reporting results, to make calculations simpler or computationally faster, or to ensure statistical accuracy.  Refs.~\cite{blennow1}, \cite{blennow2}, \cite{ciuffoli}, and \cite{qian} serve as further introduction to these issues.  We take a pragmatic view here, noting that on the whole these concerns tend not to lead to actionable differences in sensitivity estimates.  These differences are quantified within the frequentist realm in~\cite{blennow1}.  Since frequentist and Bayesian techniques answer fundamentally different questions, cross-comparisons between them are more fraught, especially at low sensitivities.  Nevertheless, the primary concerns that arise when examining sensitivity projections are rarely these fraction-of-a-$\sigma$ effects but rather the broader issues of engineering or detector performance requirements, the treatment of systematic uncertainties, the realism of stated timelines, or the prospects for obtaining funding for the desired experimental scope.

Another popular statistical question surrounding the mass hierarchy is:\ With what confidence must we establish it?  In practice, each increment in hierarchy sensitivity adds importantly to our understanding.  Even a $2\sigma$ measurement would begin guiding the model-building community and would become a meaningful input in other areas of physics ({\em e.g.},\ cosmology and astrophysics).  As the hierarchy significance increases, so too does the significance of any measurement that takes the hierarchy as input.  Thus, while a $3\sigma$ hierarchy measurement will be a historic milestone for particle physics, a continued push to $5\sigma$ or beyond is needed if we are to make $\mathord{\gtrsim}5\sigma$ claims for other, quantitatively related questions ({\em e.g.},\ exclusion of Majorana neutrinos due to future non-observation of neutrinoless double beta decay).  Whether those subsequent questions require high levels of significance is beyond the scope of this article.

\section{EXPERIMENTS}
\subsection{Long-baseline}\label{sec:lbl}
Long-baseline neutrino experiments use accelerator-based sources of $\nu_\mu$ or $\overline{\nu}_\mu$ and large detectors some hundreds of kilometers downstream to investigate neutrino oscillations.  These experiments probe the mass hierarchy through the matter effects that modify the appearance probability $P(\nu_{\mu}\ra\nu_e)$ and its antineutrino counterpart from their vacuum values.  Neglecting higher-order terms in $\alpha\equiv\Delta m^2_{21}/\Delta m^2_{32}\approx 0.03$ and $\sin^2(\theta_{13})\approx 0.02$, the $\nu_{\mu}\ra\nu_e$ transition probability for long-baseline experiments is given by~\cite{numunuematter}
\begin{equation}\label{equ:numunuematter}
\begin{split}
P(\nu_{\mu}\ra\nu_e)\ \approx\ 
    &\sin^2\theta_{23}\sin^22\theta_{13}\frac{\sin^2(\Delta(1-x))}{(1-x)^2}\\
    &+\alpha J \cos(\Delta\pm\delta)\frac{\sin(\Delta x)\sin(\Delta(1-x))}{x(1-x)}\\
    &+\alpha^2\cos^2\theta_{23}\sin^22\theta_{12}\frac{\sin^2(\Delta x)}{x^2}\ ,
\end{split}
\end{equation}
where $J\equiv\cos\theta_{13}\sin2\theta_{13}\sin2\theta_{12}\sin2\theta_{23}$, $\Delta\equiv\Delta m^2_{32}L/(4E)$, $x\equiv\pm2\sqrt{2}G_Fn_eE/\Delta m^2_{32}$, and the plus (minus) signs apply to neutrino (antineutrino) oscillations.  In the definition of $x$, $G_F$ is the weak coupling constant and $n_e$ is the number density of electrons in the propagation medium.

The mass hierarchy enters via the $(1-x)$ factors in Eq.~(\ref{equ:numunuematter}), with $x$ switching signs between the normal and inverted cases.  To increase the scale of this perturbation, one must increase the neutrino energy $E$.  However, one must also stay near the oscillation maximum of $\Delta\approx\pi/2$ to have significant oscillation probability at all.  Thus, the experimental baseline $L$ must also increase, leading to the practical fact that an increased sensitivity to the mass hierarchy in the $\nu_e$ and \anue{} appearance channels comes from having a longer experimental baseline.

\begin{figure}
  \begin{minipage}{0.49\linewidth}
    \includegraphics[width=\linewidth]{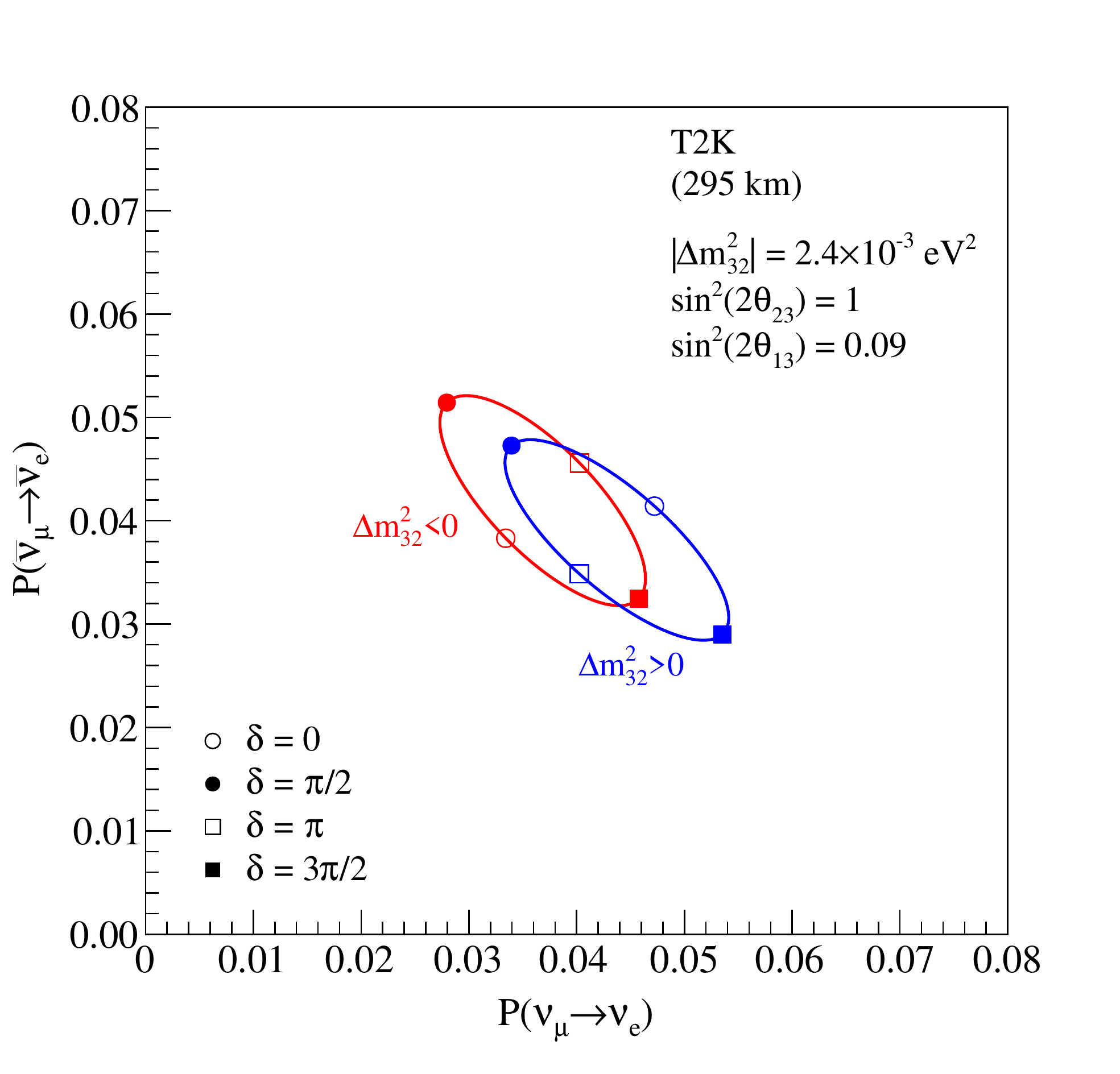}
  \end{minipage}
  \begin{minipage}{0.49\linewidth}
    \includegraphics[width=\linewidth]{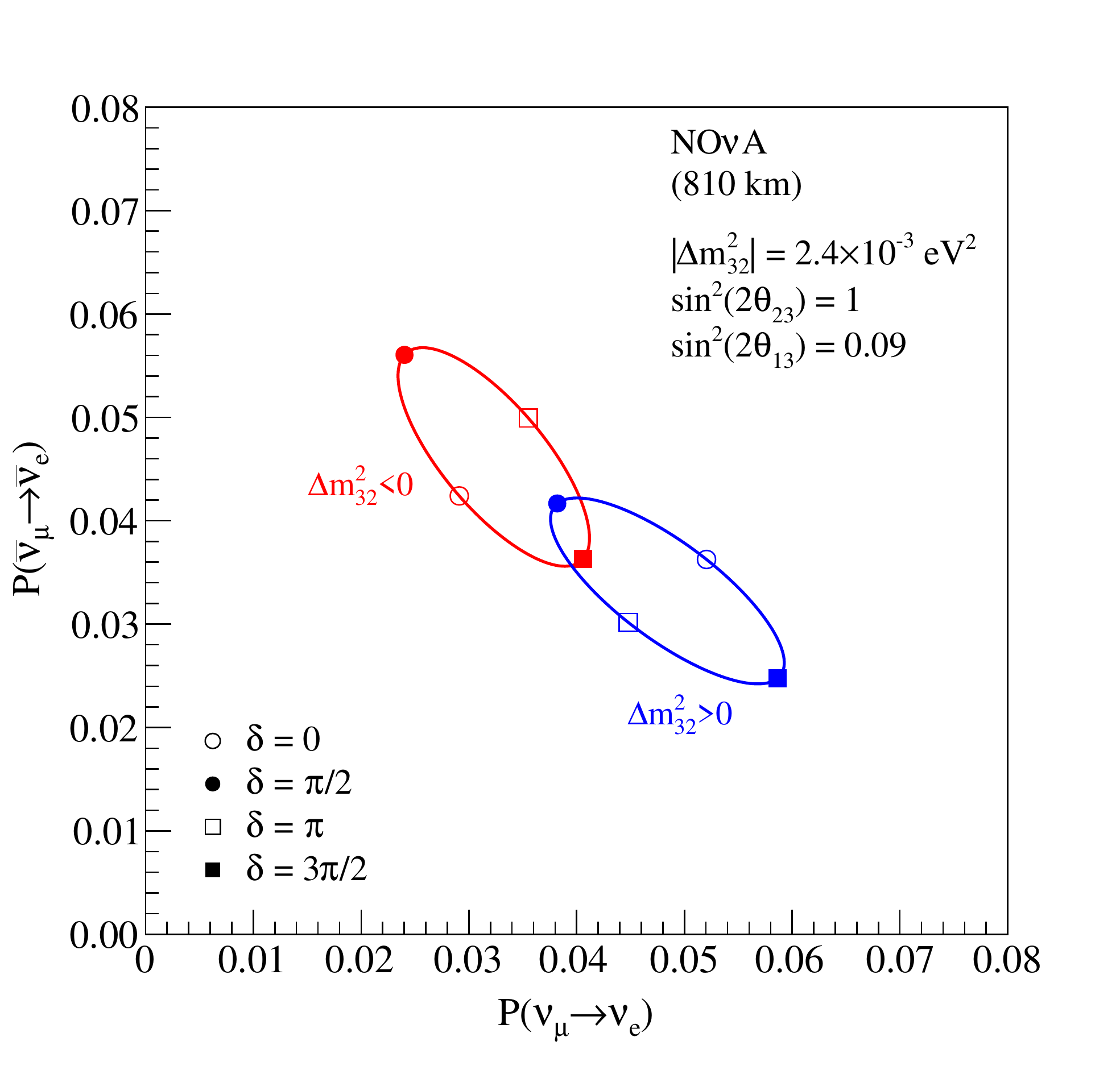}
  \end{minipage}
  \begin{minipage}{0.49\linewidth}
    \includegraphics[width=\linewidth]{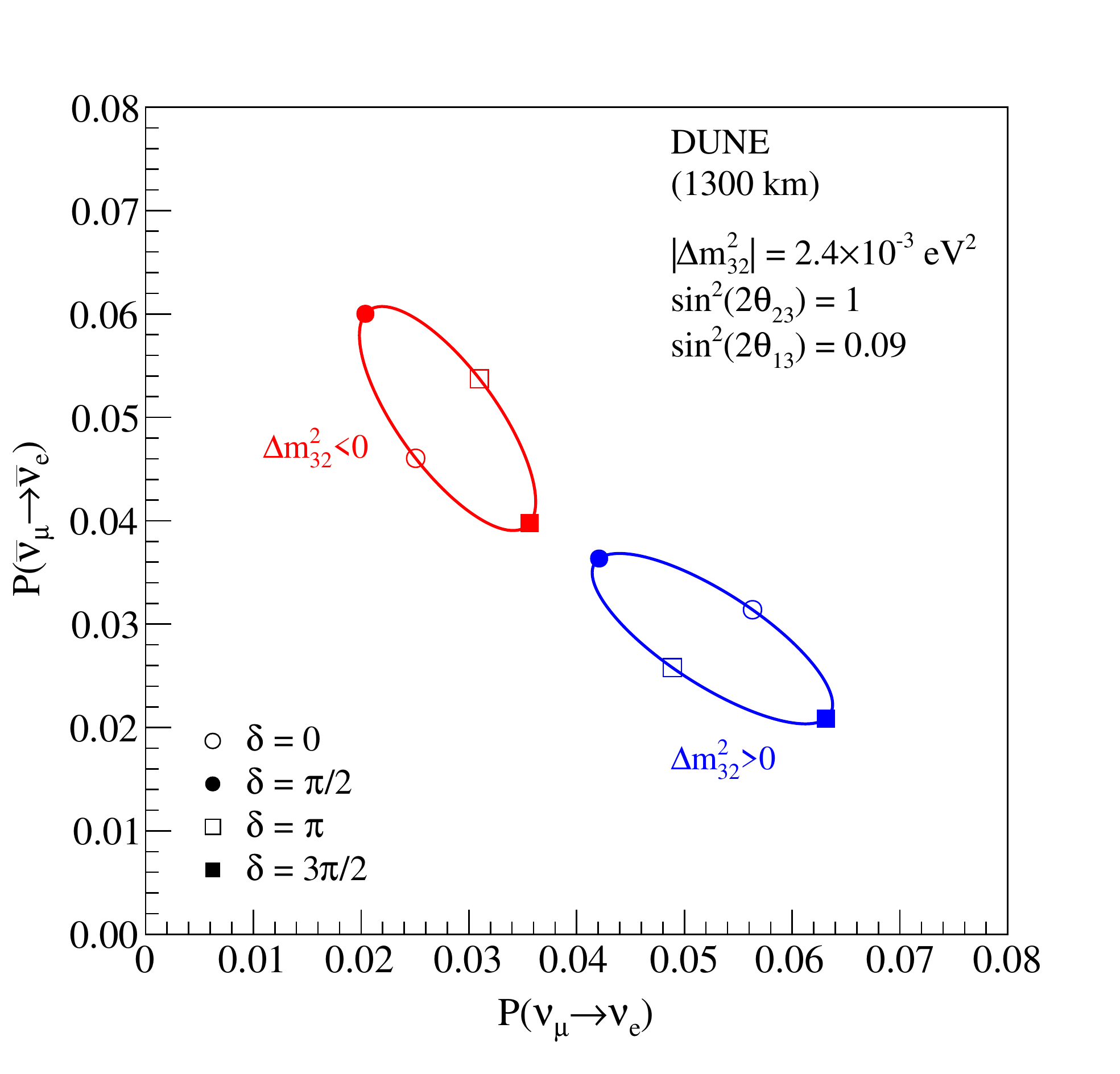}
  \end{minipage}
  \begin{minipage}{0.49\linewidth}
    \caption{$P(\overline{\nu}_\mu\ra\overline{\nu}_e)$ versus $P(\nu_\mu\ra\nu_e)$ for both hierarchies (red and blue ellipses) and for the full range of $\delta$ (cycling around the ellipses) for a representative $L/E$ value of $0.4~\mathrm{km}/\mathrm{MeV}$.  At the T2K baseline, the neutrino and antineutrino oscillation probabilities differ very little between the two hierarchies and thus measurements of these probabilities offer minimal hierarchy discrimination.  For \nova{}, significant splitting of the two cases occurs.  For DUNE, the separation is complete, allowing for unambiguous determination of the hierarchy regardless of $\delta$, assuming small enough measurement errors on the probabilities.  This figure is illustrative only, as it keeps other oscillation parameters fixed and as a full experiment involves a range of neutrino energies.}\label{fig:biprob}
  \end{minipage}
\end{figure}

Figure~\ref{fig:biprob} shows how the appearance probability changes over the three baselines discussed here:\ 295~km (T2K), 810~km (\nova{}), and 1300~km (DUNE).  The probabilities are plotted for all possible values of the {\em CP}-violating phase $\delta$, tracing out ellipses in the figure.  The uncertainty in $\delta$ complicates the mass hierarchy measurement, although less so at the longest baselines since the hierarchy-dependent matter effects become larger than any possible $\delta$-induced changes in the oscillation probabilities.  Hierarchy sensitivities for long-baseline experiments are often quoted as a function of the true (if unknown) $\delta$.  The uncertainty on $\theta_{23}$ is the next largest factor after $\delta$, with the current allowed range of $\theta_{23}$ leading to a peak sensitivity uncertainty of roughly $20$\%.  This dependence is not always explicit in experiments' official sensitivity plots, which often assume a fixed typical value for $\theta_{23}$, say $\pi/4$.

T2K and \nova{} are currently operational, and DUNE is a planned future experiment.  Several concepts for a future long-baseline experiment, some with impressive hierarchy sensitivity, have been studied and pursued in recent years.  However, most of these have remained virtual given the need for significant funding (approximately US\$1B) and a firm programmatic commitment from a suitable facility.  Today, two efforts have significant momentum.  One is DUNE, a newly formed international collaboration that, in practice, merges several previously independent efforts.  The other, Hyper-K, is discussed in Section~\ref{sec:atmos}.

\subsubsection{T2K and \nova{}}
T2K~\cite{t2k} sends neutrinos from J-PARC to the Super-Kamiokande (Super-K) detector in the Kamioka mine 295 km away.  The storied Super-K detector~\cite{superkdetector} is a 50-kton (22-kton fiducial) water Cherenkov detector situated 44 mrad off the T2K neutrino beam axis.  An off-axis arrangement, also used by \nova{}, yields a narrower neutrino energy spectrum with a reduced high-energy tail, thus decreasing the rate of neutral current events that form the dominant background for the signal $\nu_e$ charged current channel.  In 2013, T2K made the first definitive observation of $\nu_\mu\ra\nu_e$ oscillations at $7.3\sigma$ using $\sim$10\% of the experiment's eventual planned exposure~\cite{t2knue}.  However, as Figure~\ref{fig:biprob} suggests, the relatively short baseline of T2K precludes determining the mass hierarchy from this $\nu_e$ appearance data alone.

\begin{figure}
\begin{center}
  \begin{minipage}{0.48\linewidth}
    \includegraphics[width=\linewidth]{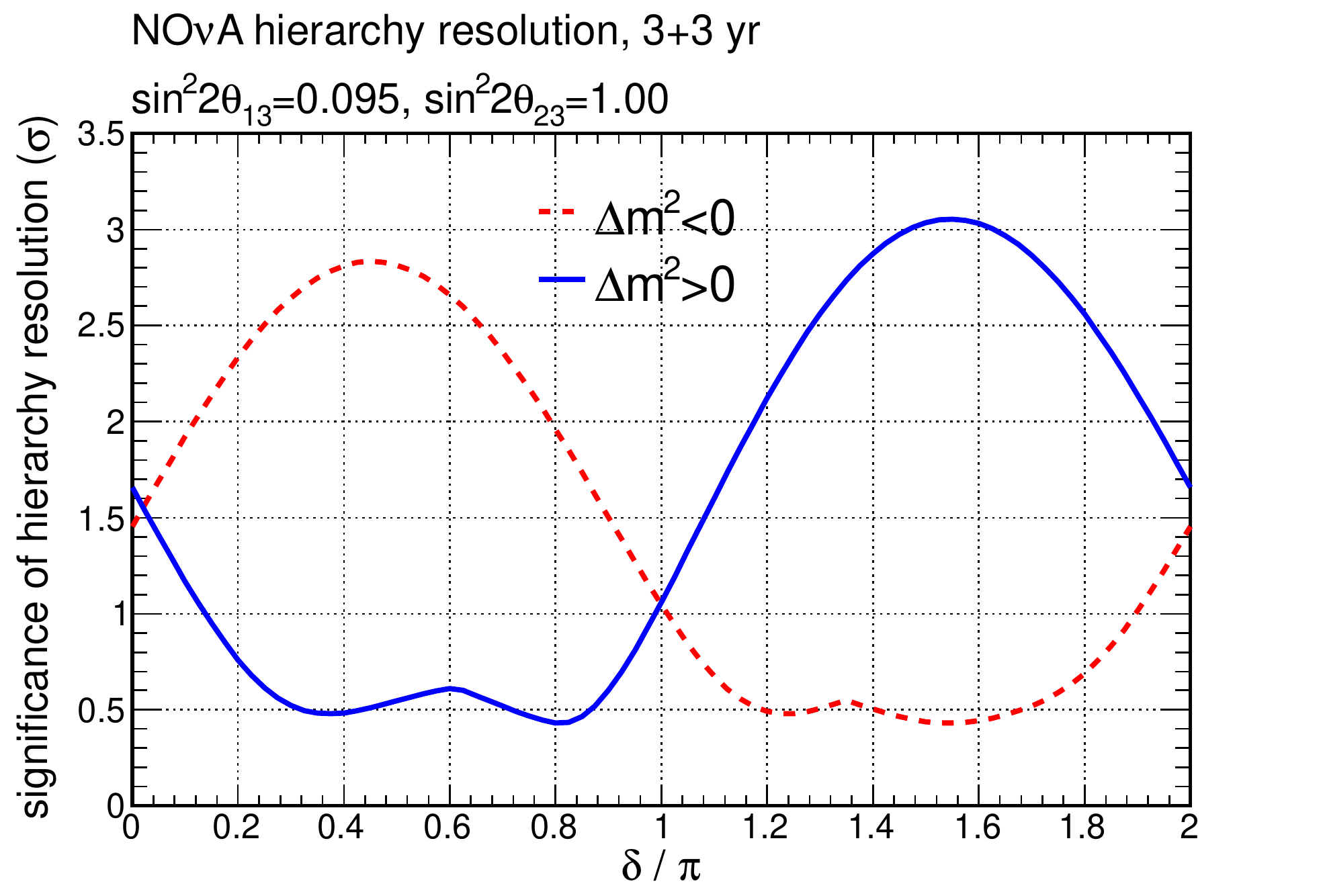}
  \end{minipage}
  \begin{minipage}{0.48\linewidth}
    \includegraphics[width=\linewidth]{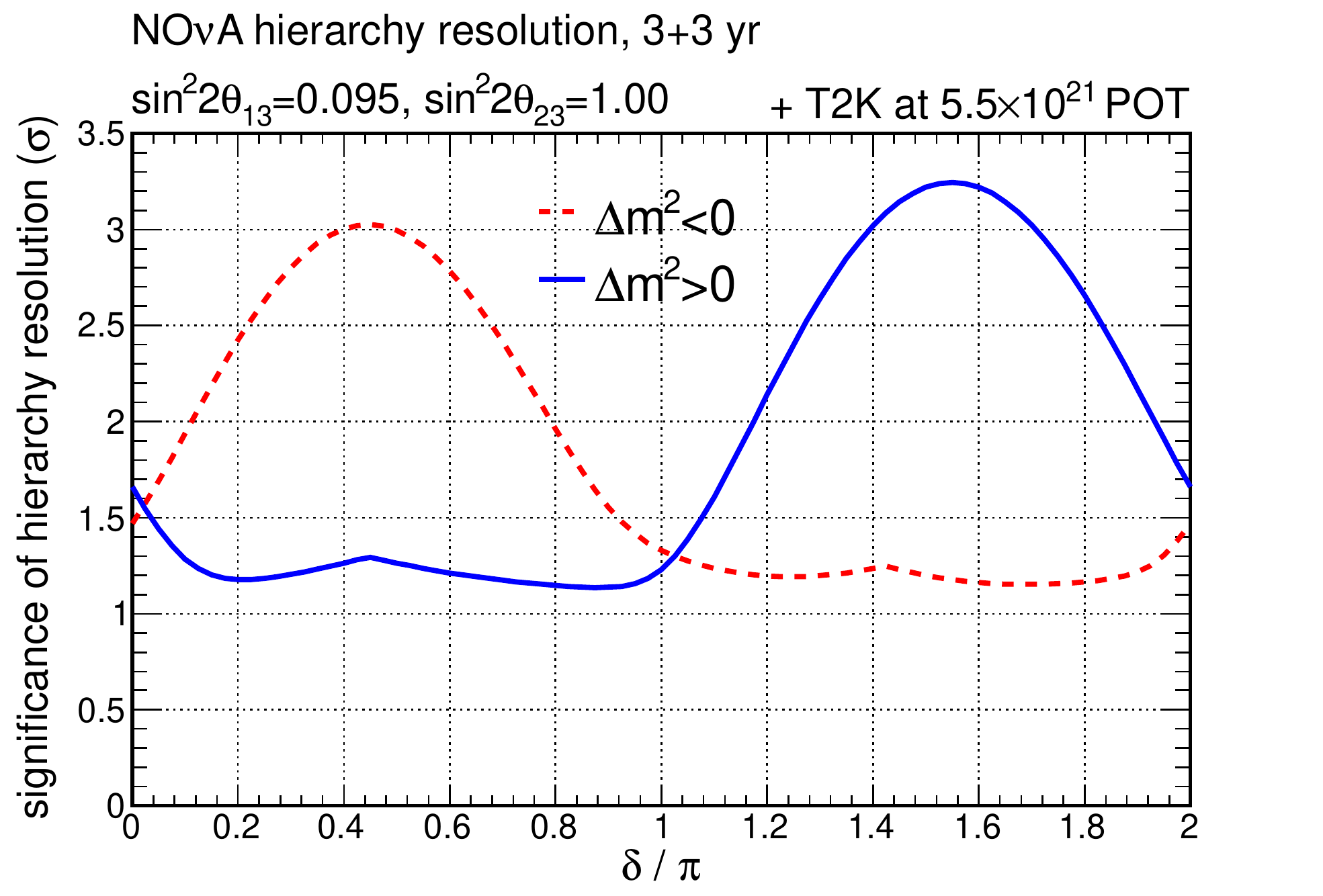}
  \end{minipage}
  \caption{({\em Left}) Hierarchy sensitivity for the planned \nova{} exposure of $3.6\mathord{\times}10^{21}$~protons on target (POT), plotted as a function of true $\delta$.  ({\em Right}) The same with the addition of $5.5\mathord{\times}10^{21}$~POT of T2K $\nu_\mu\ra\nu_e$ data.  The T2K data help break the $\delta$-vs.-hierarchy ambiguity somewhat, although outright hierarchy determination is still unlikely in the unfavorable regions of parameter space.  Figure adapted from Ref.~\cite{novasens}.}
  \label{fig:novasens}
\end{center}
\end{figure}

\nova{}~\cite{nova} uses the NuMI neutrino source at Fermilab and a new 14-kton highly segmented tracking calorimeter 810 km away and 14 mrad off the neutrino beam axis in Ash River, Minnesota.  \nova{} began taking data with a fully deployed 14-kton detector in 2014.  With its full planned exposure, \nova{} can determine the hierarchy at $2\text{--}3\sigma$ if the value of $\delta$ falls in a favorable range.  For unfavorable $\delta$ values, \nova{} extracts correlated information about $\delta$ and the hierarchy and thus does not determine the hierarchy outright.  In these so-called ``degenerate'' cases, it helps a bit to combine \nova{} and T2K data since T2K is sensitive to $\delta$ but largely insensitive to the hierarchy.  Figure~\ref{fig:novasens} shows the projected reach of \nova{} both with and without the inclusion of T2K data.  \nova{} is the only operational experiment with any significant hierarchy sensitivity, albeit with a possibility of confusion from $\delta$.

Both T2K and \nova{} use multiple detectors to mitigate systematic uncertainties.  \nova{} uses identical detector technologies at its two sites at $L\,\mathord{\approx}\,1$~km and 810~km.  T2K uses a multipurpose, off-axis near detector at 280 meters together with a beam monitoring detector situated on-axis.  The multi-detector approach makes the T2K and \nova{} experiments statistics limited for some time, so our knowledge of the hierarchy in the next five to ten years will be driven in part by the beam power achieved at Fermilab and, to a lesser extent, at J-PARC.  The NuMI beam reached 400~kW in early 2015 and is projected to reach the design 700~kW within two years, after the completion of upgrades to the Booster, which lies upstream in the accelerator chain.  The J-PARC neutrino source is operating at $\sim$325~kW, and significant future upgrades are needed to realize the planned 750~kW.

\subsubsection{DUNE}\label{sec:DUNE}
In the past decade, significant global effort has gone into designing a next-generation long-baseline experiment, and well-developed proposals have come out of this process.  The most challenging aspect at present is budgetary.  The full physics scope of a future experiment -- precision measurements of PMNS matrix elements, {\em CP} violation, proton decay, supernova physics, and beyond-the-Standard-Model searches -- requires a substantial detector and a beamline to match, at a combined cost surpassing US\$1B.  Such a program is only achievable through international partnerships, and a new collaboration, originally named ELBNF and since renames DUNE, has recently formed with the early goal of carrying forward the significant design work already completed to form a fully international long-baseline experimental proposal~\cite{dune}.

Fermilab will host the planned 1.2-MW neutrino source for DUNE, and the experiment's far detector will be located in the Sanford Underground Research Facility (SURF) in South Dakota, yielding a baseline of 1,300 km.  The far detector will be a liquid argon time-projection chamber (TPC) with a total mass of 40~kton and a fiducial mass around 34~kton.  One of the existing design efforts, LBNE, also assumed a Fermilab-to-SURF layout with a large liquid argon detector, so we use the LBNE sensitivity estimates here~\cite{lbne} as a proxy for DUNE.

\begin{figure}
\begin{center}
  \includegraphics[width=0.55\linewidth]{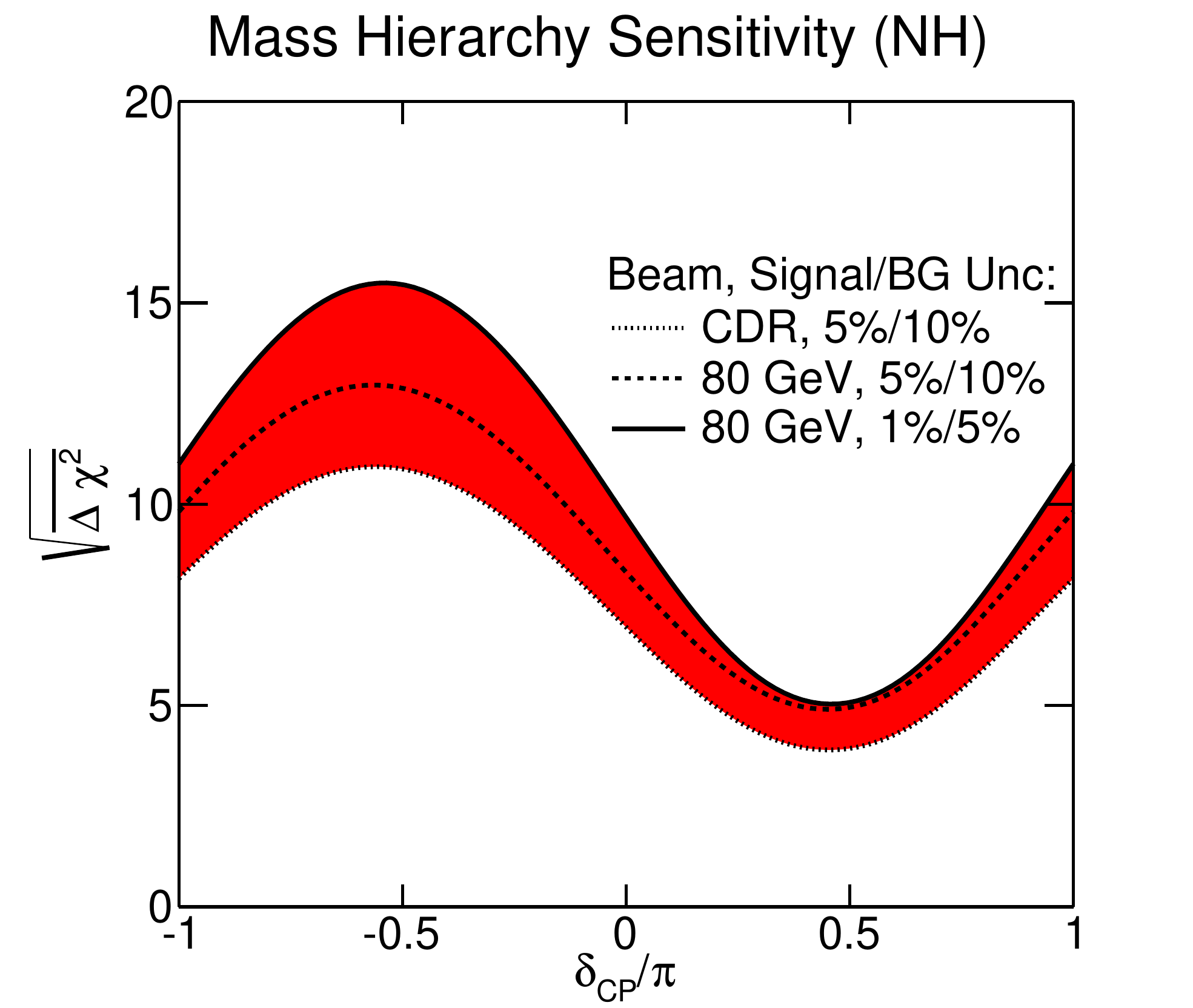}
  \caption{Hierarchy sensitivity for DUNE at a fiducial exposure of (1.2 MW)$\times$(34 kton)$\times$(10 yr).  The red band spans the range of assumptions shown:\ ({\em dotted}) beam design in the LBNE conceptual design report and 5\% (10\%) uncertainties taken for the signal (background), ({\em dashed}) a further optimized beam design using 80~GeV protons, and ({\em solid}) the same as the dashed line but with more aggressive systematic uncertainties.  Normal hierarchy is assumed.  Figure adapted from Ref.~\cite{lbne}.}
  \label{fig:dune}
\end{center}
\end{figure}

The ten-year hierarchy sensitivity for a 40-kton DUNE is substantial (Figure~\ref{fig:dune}).  Even at the least favorable $\delta$ and $\theta_{23}$ values, DUNE achieves $5\sigma$ significance and exceeds $10\sigma$ for other parameter assumptions.  Although some open questions remain regarding near detector performance, far detector event selections, and systematic uncertainties, these hierarchy sensitivity estimates are unlikely to vary grossly as the open questions are addressed.  The greatest uncertainty in the reach of DUNE is in the timeline and scope of the experiment.  The collaboration aims to deploy an initial 10-kton detector underground at SURF by 2021, followed shortly by completion of the neutrino source and further deployment to 40 kton.  These timescales are achievable but aggressive.  The recent US\ high energy physics strategic plan~\cite{p5} strongly supports this effort, and operation of a 10-kton detector and a 1.2-MW source is estimated to begin in 2025.  How quickly additional detector mass can be added and how quickly a 1.2-MW beam can be deployed will play a large role in the sensitivities realized.  To good approximation, the sensitivities of Figure~\ref{fig:dune} can be scaled according to the square root of exposure.

\subsection{Atmospheric}\label{sec:atmos}
Atmospheric neutrino experiments take advantage of the flux of neutrinos -- a reasonably well-characterized mix of \numu{}, \anumu{}, \nue{}, and \anue{} -- produced by cosmic ray showers in the atmosphere.  Since detectors can collect these neutrinos from all directions, the distance between neutrino production and detection can range from tens of kilometers for neutrinos produced above the detector to 13,000 kilometers for neutrinos produced below ({\em i.e.},\ on the other side of the Earth).  This wide range of baselines, together with a wide range of energies, makes atmospheric neutrinos particularly useful probes of neutrino oscillations.

\begin{figure}[t]
\begin{center}
  \begin{minipage}{0.48\linewidth}
    \includegraphics[width=\linewidth,viewport=15 0 391 306,clip=true]{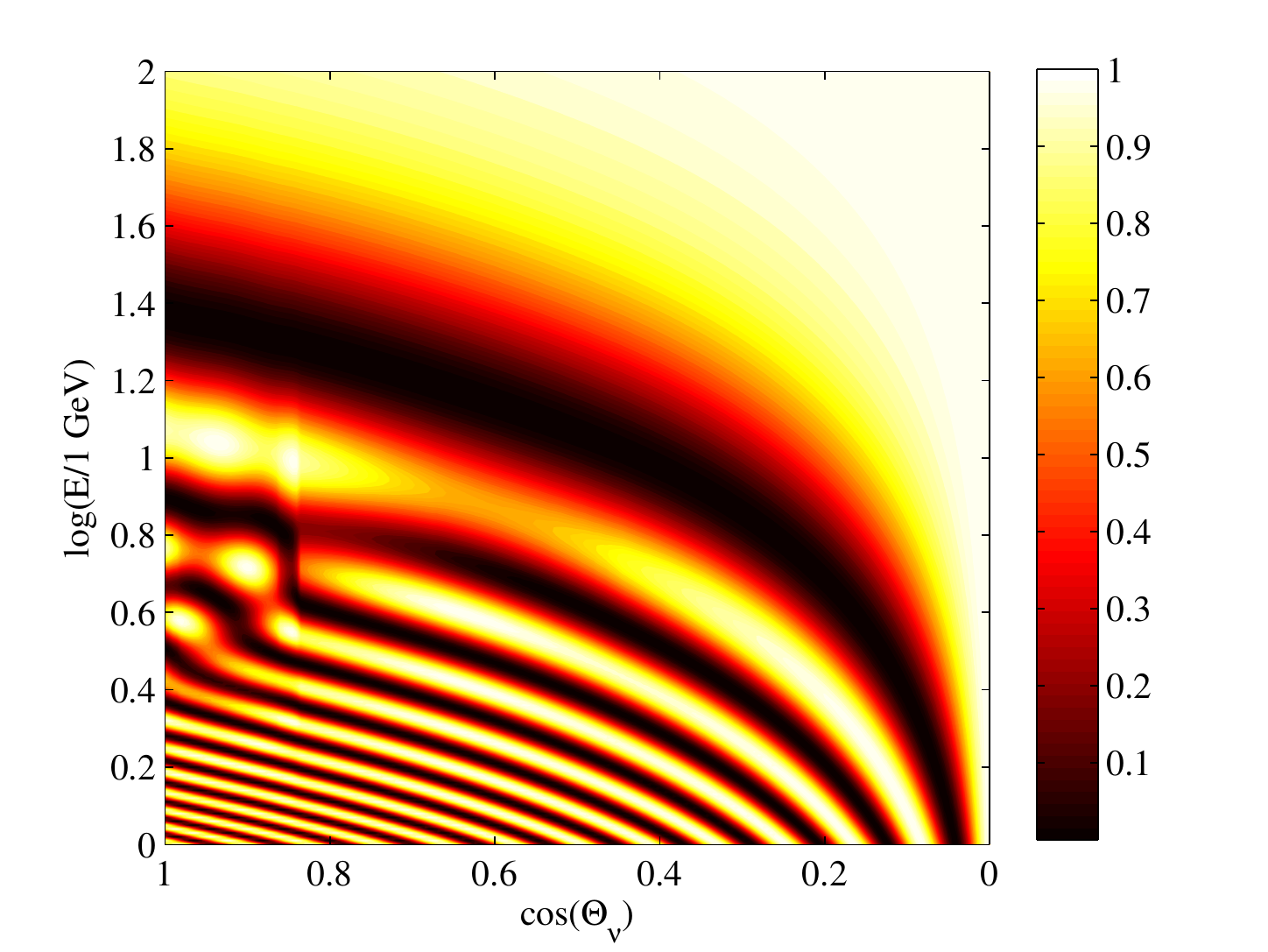}
  \end{minipage}
  \begin{minipage}{0.48\linewidth}
    \includegraphics[width=\linewidth,viewport=15 0 391 306,clip=true]{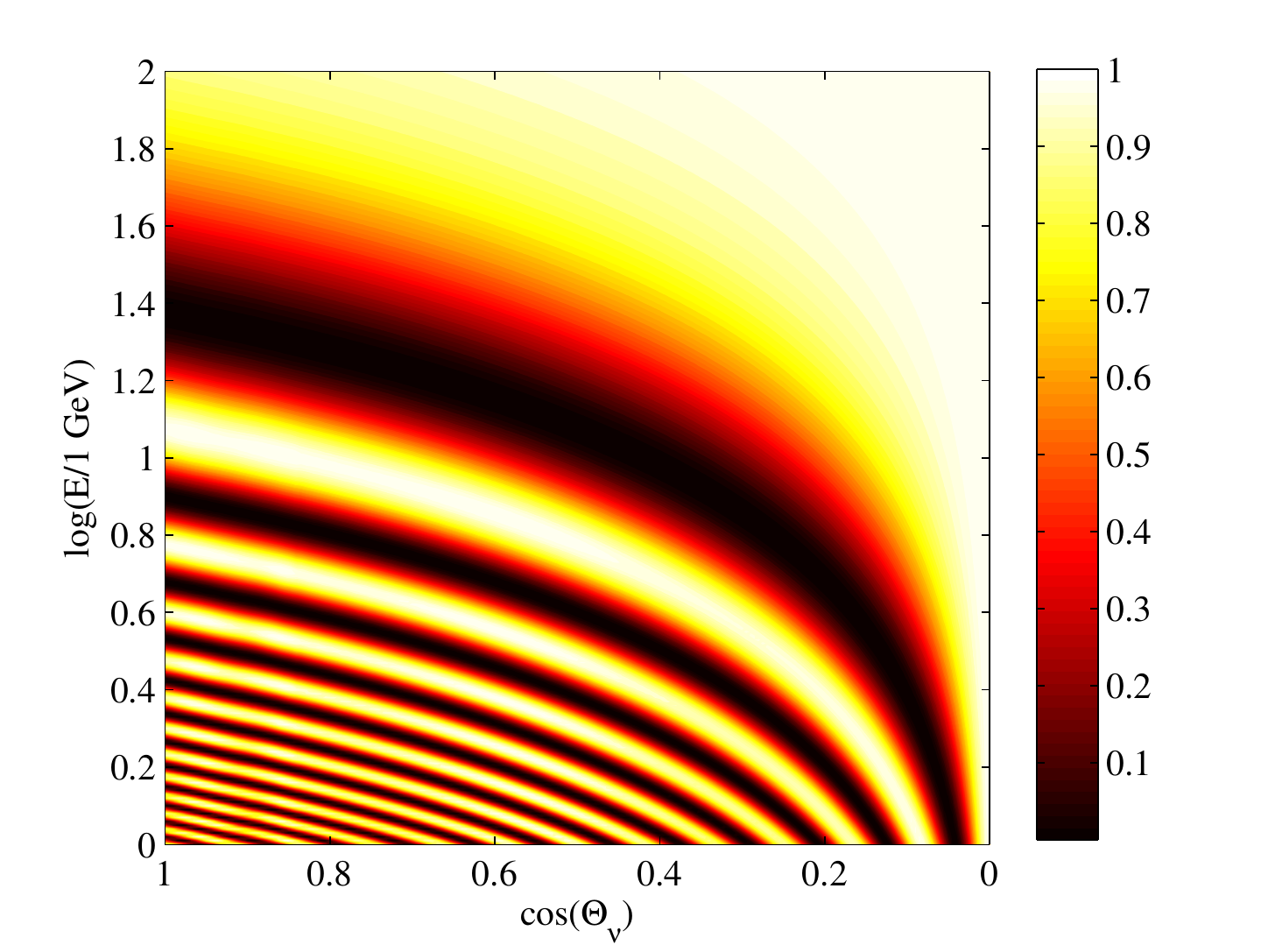}
  \end{minipage}
  \caption{Oscillation probabilities $P(\numu\ra\numu)$ ({\em left}) and $P(\anumu\ra\anumu)$ ({\em right}) for neutrinos passing through the Earth for the normal mass hierarchy.  The probabilities are plotted as a function of the neutrino energy $E$ (expressed as $\log_{10}(E/\mathrm{GeV})$) and the direction of travel $\Theta_\nu$, with $\cos(\Theta_\nu)\mathord{=}1$ corresponding to an upward neutrino trajectory into the detector.  These neutrinos will have passed through the full diameter of the Earth.  Trajectories at $\cos(\Theta_\nu)\mathord{=}0$ correspond to the horizon.  The matter effect enhancement is seen in $P(\numu\ra\numu)$ at $2~\mathrm{GeV}<E<15~\mathrm{GeV}$ and is absent in $P(\anumu\ra\anumu)$.  The situation is reversed for the inverted hierarchy.  A corresponding effect occurs for the \nue{} and \anue{} appearance channels.  Figure adapted from Ref.~\cite{earthmatter3}.}
  \label{fig:oscillogram}
\end{center}
\end{figure}

Atmospheric neutrino experiments gain hierarchy sensitivity through the resonant-like enhancement of matter effects experienced by either neutrinos or antineutrinos (depending on the hierarchy) as they pass through the Earth~\cite{earthmatter1,earthmatter2,earthmatter3}.  The effect on the oscillation probabilities is often presented in so-called oscillograms ({\em e.g.},  Figure~\ref{fig:oscillogram}).  Experiments that can study neutrinos and antineutrinos separately need only determine which sample exhibits the matter effect enhancement.  Experiments that cannot separate these must make a statistical inference based on knowledge about the incident neutrino and antineutrino fluxes.  Current experiments, notably Super-K~\cite{superkatm} and MINOS~\cite{minos3nu}, have demonstrated the principle of this measurement, but these detectors are too small and thus have too few events to obtain significant hierarchy sensitivity, despite their decade-long exposures.  Proposed future large detectors fall into three categories:\ water Cherenkov (PINGU, ORCA, Hyper-K), magnetized iron tracker (INO), and liquid argon TPC (DUNE).

\subsubsection{PINGU, ORCA, and Hyper-K}
PINGU (Precision IceCube Next Generation Upgrade)~\cite{pingu} is a proposed extension to the IceCube detector~\cite{icecube} at the South Pole.  PINGU would add an array of 40 strings each with 60 optical modules to the DeepCore region of IceCube, spanning 4 Mton of ice.  As the technology is well established, the cost and schedule estimates are robust, and operation could begin by 2020 if funding is established soon.  The relatively high density of optical modules in PINGU versus the rest of IceCube would lower the triggering threshold to $\sim$1~GeV and would provide good energy and angle resolutions for events in the $5\text{--}15\ \mathrm{GeV}$ region in which the hierarchy signature lives.  These resolutions are critical inputs to the sensitivity estimates, as the matter effects of Figure~\ref{fig:oscillogram} are heavily smeared.

\begin{figure}[t]
\begin{center}
  \includegraphics[width=0.65\linewidth]{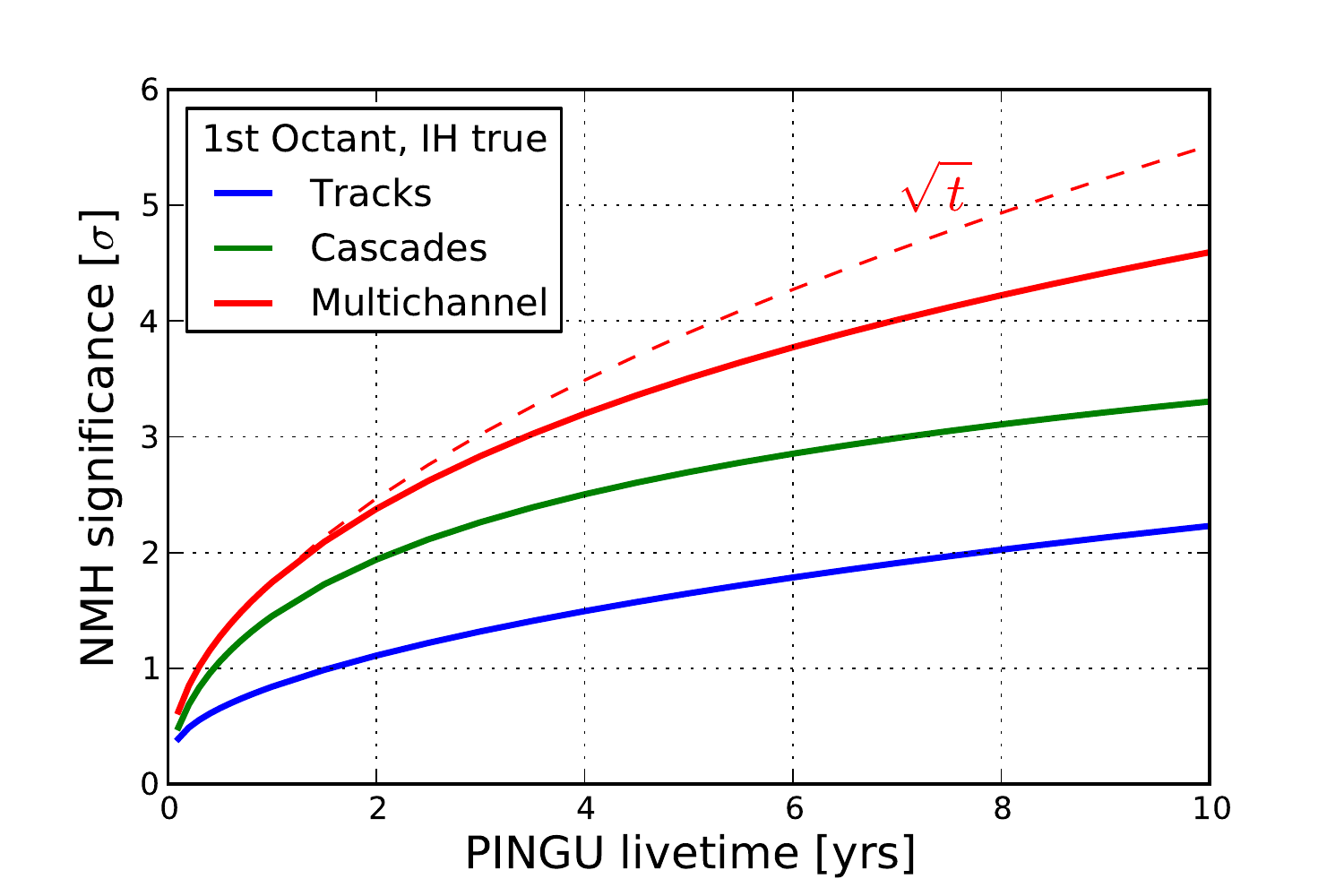}
  \caption{Mass hierarchy sensitivity for PINGU as a function of livetime.  The solid red ``multichannel'' curve represents the best estimate.  In the past, atmospheric experiments focused on the track events from \numu{} and \anumu{} charged current interactions, which on their own lead to the bottom blue curve (``tracks'').  More recently the power of the cascade events, particularly \nue{} and \anue{} charged current interactions, has been taken into consideration even for detectors with relatively poor particle identification capabilities.  The deviation of the multichannel curve from a $\sqrt{t}$ behavior (dashed red) provides a measure of the impact of systematic uncertainties.  Figure adapted from Ref.~\cite{pingu}.}
  \label{fig:pingu}
\end{center}
\end{figure}

PINGU projections have evolved significantly in the past couple of years.  Detailed studies of detector performance and the effects of systematic uncertainties (notably energy scale; $\nu$ and $\overline{\nu}$ cross sections; and the oscillation parameters $\Delta m^2_{32}$, $\theta_{23}$, and $\theta_{13}$) have been carried out~\cite{pingu}.  Further, the prospects for improvements through detector geometry optimization, incorporation of event inelasticity as a weak neutrino/antineutrino discriminant~\cite{inelasticity}, and more advanced reconstruction techniques suggest that the current sensitivity estimates may be conservative.  Figure~\ref{fig:pingu} shows the sensitivity of the combined track ($\nu_\mu$ charged current) and cascade (all other) channels in PINGU, which reaches $3\sigma$ hierarchy sensitivity in 4 years of running if $\theta_{23}$ is in the lower octant.

ORCA (Oscillation Research with Cosmics in the Abyss)~\cite{orca} is a proposed option for a multi-megaton water Cherenkov array based on the KM3NeT~\cite{km3net} deep-sea technology.  The measurement strategy is the same as for PINGU:\ use a dense array of optical modules to obtain adequate energy resolution, angle resolution, and event identification capabilities to observe the matter effect enhancement for neutrinos below 15~GeV.  The latest ORCA sensitivity studies have included track and cascade channels, and the proposed detector has grown from 2 Mton with 50 strings to 4 Mton with 115 strings.  The stated ORCA sensitivities are on par with, or somewhat better than, those of PINGU, with $3\sigma$ sensitivity anticipated in 3 years~\cite{nextgenatmos}.  However, the ORCA estimates are less well developed and documented than those of PINGU, so it is difficult to assess how they will evolve as work on systematic uncertainties and analysis techniques continues.

The proposed Hyper-K detector~\cite{hyperk} consists of two cylindrical water-filled tanks with a total mass of 1~Mton and with 20\% photocathode coverage provided by 10,000 photomultiplier tubes.  Hyper-K would be sited near the Kamioka mine and would also serve as the far detector for the long-baseline T2HK experiment, a scaled-up version of T2K.  Like T2K, T2HK has limited hierarchy sensitivity on its own given its baseline, and thus Hyper-K's hierarchy sensitivity comes primarily from atmospheric neutrinos.  The excellent event reconstruction capabilities of the Hyper-K detector are well understood based on long Super-K experience.  The most glaring uncertainty for Hyper-K is whether the project proceeds.  Whereas PINGU and ORCA are relatively low-cost options (US\$50M--\$80M), Hyper-K would require US\$500M--\$700M over a seven-year construction timeline.  Given the focus of the U.S.\ and European long-baseline communities on liquid argon TPC approaches, substantial funding may need to come from Japan alone.  This higher cost over PINGU and ORCA, of course, brings with it a rich physics program including leptonic {\em CP} violation, proton decay, and a range of solar and astrophysical measurements~\cite{hyperk}.

Hyper-K can reach $3\sigma$ mass hierarchy sensitivity after ten years of operation assuming a lower-octant value for $\theta_{23}$ of 0.4.  This significance increases to $6\sigma$ (or, equivalently, the time to $3\sigma$ falls below three years) for an upper-octant value of $\theta_{23}\,\mathord{=}\,0.6$.  This strong dependence of hierarchy reach on $\theta_{23}$ is characteristic of all atmospheric experiments.  The sensitivities for PINGU and ORCA quoted above take conservative values of $\theta_{23}\,\mathord{\approx}\,0.4$.  For $\theta_{23}\,\mathord{=}\,0.6$, PINGU can reach $3\sigma$ sensitivity within its first year of operation and $6\sigma$ after four years.

\subsubsection{ICAL}
The ICAL detector at the India-based Neutrino Observatory (ICAL@INO)~\cite{ino} is a planned 50-kton magnetized tracker made of 150 alternating layers of iron and resistive plate chambers.  The 1.5-T magnetic field provides excellent $\mu^-/\mu^+$ separation, and consequently $\nu_\mu/\overline{\nu}_\mu$ separation, on an event-by-event basis.  This advantage, together with the detector's excellent energy and angular resolution, mitigates to some degree the lower event rates due to ICAL's small fiducial mass.  The ICAL hierarchy sensitivity after ten years of operation ranges from $2.5\sigma$ to $3.5\sigma$ depending on the true value of $\theta_{23}$.

ICAL@INO is moving ahead with support from India, and site work has begun.  Operations are projected to begin in 2018, although given the required civil construction that date is likely optimistic.  The detector technology is well-established and should not present problems scaling up to 50 kton.  Continuing detector R\&D is focused on performance and longevity optimization~\cite{inonim}.  One concern is that the sensitivity is somewhat low compared with that of other experiments proposing results on a similar time frame ({\em c.}~2030).

\subsubsection{Large liquid argon TPC}
In contrast to Hyper-K, for which the atmospheric neutrino sample provides the bulk of the hierarchy sensitivity, DUNE's atmospheric measurement is secondary to its accelerator-based measurement (Section~\ref{sec:DUNE}).  A 340-kton-year atmospheric exposure yields $3\text{--}5\sigma$ sensitivity, depending on $\theta_{23}$.  This can be compared with the accelerator-based measurement that bottoms out at $5\sigma$ in the extreme worst case.  However, if the beam power of 1.2~MW is significantly delayed relative to the detector deployment, the DUNE atmospheric sample could play a role, especially because the detector's excellent energy and angular resolution make it particularly good for this application.

\subsection{Reactor}
For the $\mathord{\sim}1\text{--}8\text{-}\mathrm{MeV}$ \anue{} streaming from a nuclear reactor, the first oscillation maximum for $\Delta m^2_{21}$-driven solar oscillations occurs at $L\mathord{\sim}100~\mathrm{km}$ and for $\Delta m^2_{32}$-driven atmospheric oscillations at $L\mathord{\sim}2~\mathrm{km}$.  KamLAND has successfully exploited the former to measure $\theta_{12}$ and $\Delta m^2_{21}$~\cite{gando}.  More recently Daya Bay, RENO, and Double Chooz have exploited the latter to measure $\theta_{13}$ and the effective squared-mass splitting $\Delta m^2_{ee}$~\cite{dayabay,reno,doublechooz}.

\begin{figure}
\begin{center}
  \includegraphics[width=0.65\linewidth]{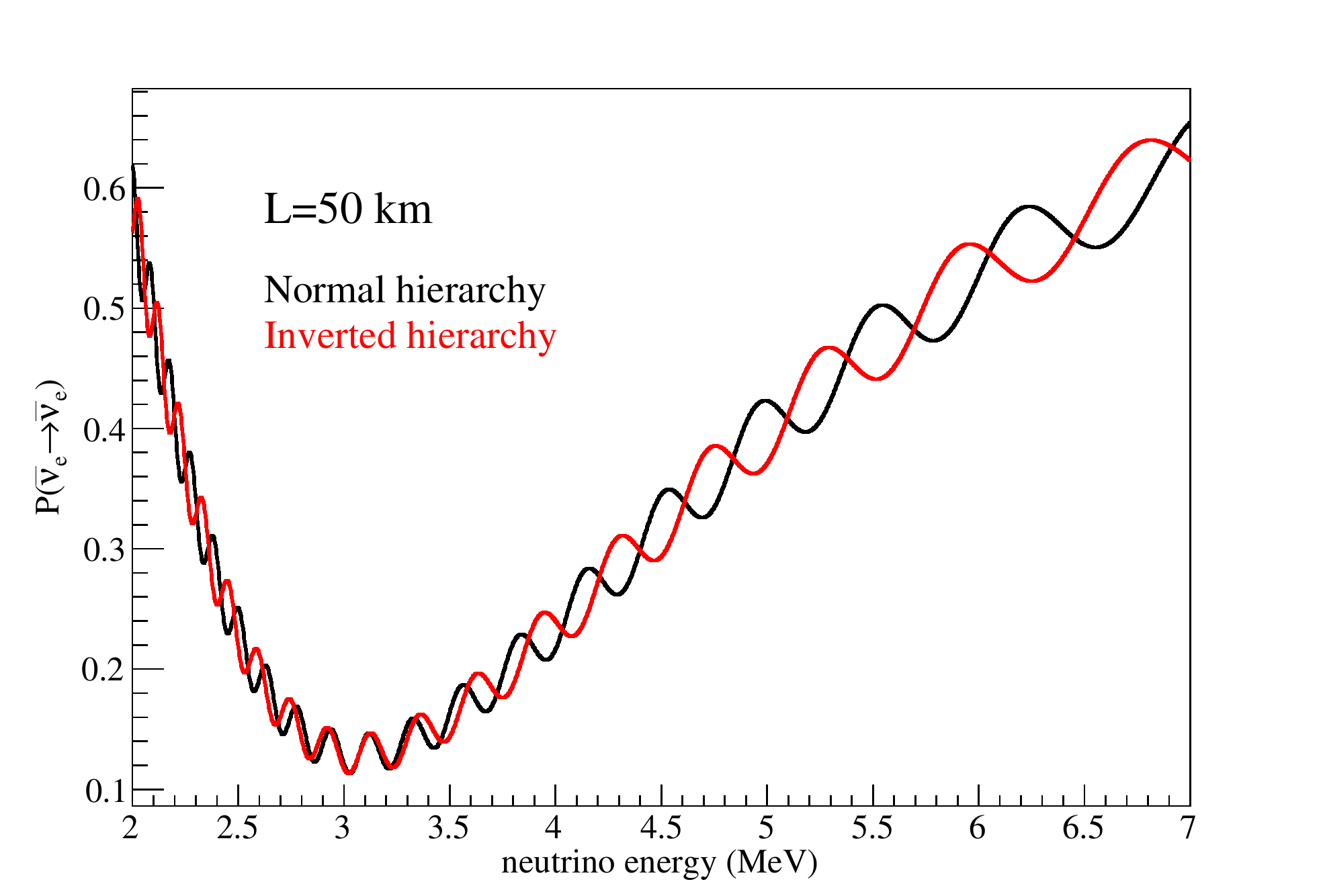}
  \caption{The oscillation probability $P(\anue\ra\anue)$ as a function of neutrino energy for a fixed neutrino baseline of 50~km.  The range of energies shown spans the relevant reactor \anue{} flux.  The wide dip centered at 3~MeV is due to the dominant solar oscillation, and the wiggles correspond to the sub-dominant atmospheric oscillation.  The two curves differ only in the assumed sign of $\Delta m^2_{32}$ (and also in the magnitude of $\Delta m^2_{31}$ since the magnitudes of the other two splittings were held fixed.)}
  \label{fig:junoprobs}
\end{center}
\end{figure}

At the longer of these baselines, the small amplitude and rapid (with respect to $E$) atmospheric oscillation will be superimposed on a deeper and slower solar oscillation.  The relative phase of these oscillation modes reveals the relative sizes of $\Delta m^2_{31}$ and $\Delta m^2_{32}$.  Figure~\ref{fig:junoprobs} demonstrates the principle of the measurement for a baseline of 50~km.  The broad solar dip in the \anue{} survival probability is visible together with the fast atmospheric oscillations, and flipping the sign of $\Delta m^2_{32}$ shifts the period and phase of these oscillations in an observable way.

Leveraging this effect requires a large, low-background detector with unprecedented energy resolution.  Over the past decade, the experimental requirements have been extensively explored~\cite{reactor1a,reactor1b,reactor2a,reactor2b,reactor2c,reactor2d,nonlinearities}, leading today to two experimental proposals, JUNO and RENO-50~\cite{juno,reno50}.  JUNO (Jiangmen Underground Neutrino Observatory) proposes a 20-kton liquid scintillator detector with $3\%/\sqrt{E/\mathrm{MeV}}$ energy resolution.  For comparison, KamLAND used a 1-kton detector with an energy resolution of $6.4\%/\sqrt{E/\mathrm{MeV}}$~\cite{gando}.  The required advance in energy resolution is to come from increased photocathode coverage, higher-efficiency phototubes, and brighter yet more transparent scintillator.  R\&D on these fronts has led to encouraging progress, but significant work remains to demonstrate that this performance is achievable~\cite{nu2014reactor}.

The pattern of the fast oscillations is also susceptible to systematic uncertainties.  An error in the magnitude of $\Delta m^2_{32}$ or in the overall energy scale of the detector will skew the oscillation pattern in a similar, though not identical, way to flipping the hierarchy.  Energy non-linearities are more pernicious, and Ref.~\cite{nonlinearities} derives a requirement of $\lesssim$0.5\% variation in energy scale across the spectrum.  This requirement can be relaxed to $\sim$2\% (a performance demonstrated with the smaller KamLAND detector) if the functional form of the non-linearity is known at some level~\cite{juno}.

JUNO is sited equidistant from two primary reactor sites, each with multiple cores.  Each core corresponds to a slightly different baseline, and the JUNO sensitivities take into account the resulting smearing of the oscillation pattern.  With a 6-year run, JUNO projects a mass hierarchy sensitivity of 4$\sigma$, give or take 10\% depending on the level of precision achieved by long-baseline experiments on the atmospheric mass splitting.  JUNO has strong support from China and is moving ahead, with operations to start around 2020.  If the detector R\&D does not lead to the performance metrics required, JUNO's hierarchy sensitivity could be considerably reduced.  However, JUNO has an impressive precision oscillation physics program separate from the mass hierarchy measurement~\cite{nu2014reactor}.

RENO-50 is a similar proposal, based in Korea, to use an 18-kton detector with performance requirements that are essentially identical to those of JUNO~\cite{reno50}.  While sensitivities of $3\sigma$ have appeared in the literature~\cite{reno50sigma}, the calculations are notably less detailed than those of JUNO ({\em e.g.},\ accounting for the smearing from multiple reactors).  Detector R\&D funding for RENO-50 has been established, but start of operations by 2020 as proposed by the experiment's proponents is likely optimistic.

\subsection{Cosmology}
Neutrinos in the early universe damped out density fluctuations on scales smaller than their free streaming length, which began decreasing as the neutrinos red-shifted toward non-relativistic energies.  The masses of the neutrinos governed the timing of the non-relativistic transition.  Thus, these masses are imprinted on the matter power spectrum today.  Multiple experimental techniques are sensitive to this imprint, including cosmic microwave background anisotropy and $B$-mode polarization measurements, galaxy surveys, weak lensing, and the Lyman-$\alpha$ forest~\cite{cosmo,cosmo2}.

Cosmological measurements are sensitive primarily to the sum of the neutrino masses, not the individual masses.  The most powerful constraints on this sum naturally comes from combinations of data sets, and current upper limits fall at $\sum m_i\lesssim0.5\ \mathrm{eV}$~\cite{pdg}.  Upcoming terrestrial and space-based measurements that can improve on these limits include MS-DESI, Euclid, LSST, and a Stage-IV CMB project.  All of these projects are relatively likely to proceed, with timescales for start of operations ranging from 2018 to 2022.  The forecast uncertainties on $\sum m_i$ begin at 20 meV ({\em c.}~2025) and improve to $\sim$10~meV ({\em c.}~2030), with past data sets taken in combination with future ones~\cite{hierllbl}.

For the normal hierarchy, $\sum m_i$ must be at least 0.06~eV given the oscillation measurements of Eqs.~(\ref{equ:solarsplit}) and (\ref{equ:atmsplit}).  For the inverted hierarchy, $\sum m_i$ must be at least 0.1~eV.  Thus if the hierarchy is normal and $m_1\,\mathord{\ll}\,m_3$, the uncertainties given above for $\sum m_i$ translate into $2\sigma$ sensitivity for determining the neutrino mass hierarchy by 2025 and $4\sigma$ a few years later.  The requirements on the mass spectrum for this measurement are critical:\ an inverted hierarchy or a smallest neutrino mass larger than $\sim$0.04~eV eliminates entirely the hierarchy-determining power of cosmological observations.

\section{OUTLOOK}
It is useful to summarize the experimental sensitivities given above in a time-ordered, rather than technique-ordered, manner.  We skip here those options with underdeveloped estimates.
\begin{itemize}
\item {\em Today:}\ Super-K (atmospheric), MINOS (atmospheric), and T2K (accelerator) have proof-of-principle results that constrain the mass hierarchy at an insignificant level.  These sensitivities will not improve substantially with added exposure.
\item {\em By 2020:}\ \nova{} will have most of its planned data set and can make a statement at $2\text{--}3\sigma$ for favorable values of $\delta$.  For unfavorable values, \nova{}'s hierarchy measurement will be correlated with its $\delta$ measurement, and an unambiguous hierarchy determination will not yet be possible.  In these cases, combining with T2K will help marginally.
\item {\em By 2025:}\ PINGU could provide $3\text{--}6\sigma$ sensitivity if the experiment is deployed in a timely manner.  If the JUNO detectors perform as required, a partial JUNO exposure could provide $2\text{--}3\sigma$ sensitivity by this date.
\item {\em By 2030:}\ Some uncertain fraction ($<$50\%) of DUNE's 340-kton-year exposure could be available, yielding sensitivities of $\mathord{\geq}3\sigma$.  Similarly, Hyper-K could provide $2\text{--}4\sigma$, and JUNO's sensitivity would reach its design $4\sigma$.  Input from cosmology would be firmly in hand.  ICAL could reach $3\sigma$ by this time.
\item {\em By 2035:}\ If DUNE proceeds in a timely manner, a large fraction of its full exposure would be available, and the hierarchy would be definitively established at $\mathord{>}5\sigma$.  Hyper-K would be at or beyond its design $3\text{--}6\sigma$.
\end{itemize}
Given the complementarity of the various techniques, it is natural to explore combined sensitivities, and the literature contains a near-exhaustive set of combinations.  The \nova{} and T2K combination is discussed in Section~\ref{sec:lbl}, above.  Atmospheric and reactor experiments offer a particularly interesting synergy, as fits to the wrong hierarchy introduce different pulls on $\Delta m^2_{32}$ between the two techniques.  The PINGU$+$JUNO case is explored in Ref.~\cite{pingujuno}.

Setting aside the potential sensitivity boosts via combinations, however, it is critical to  measure the hierarchy through independent techniques, not only for the added confidence in the results but also as protection against unfavorable parameters of nature.  To wit, atmospheric sensitivities are largely independent of $\delta$, in contrast to accelerator sensitivities.   Reactor sensitivities are invariant with respect to $\theta_{23}$, in contrast to atmospheric sensitivities.  Cosmological measurements are entirely free of dependence on oscillation parameters but do require a very particular mass spectrum.

Table~\ref{tab:summary} summarizes the experimental situation and the prospects for determining the neutrino mass hierarchy in the coming years.  The table lists the most critical concerns, some of which are major.  However, if even just a couple of the promising experimental options becomes reality without delay, the outlook can be succinctly stated:\ A statistically significant ($>$95\% CL)\ hierarchy determination is possible, but not guaranteed, by 2020; a statistically significant determination is likely by the late 2020s; and a definitive ($\mathord{>}5\sigma$) measurement is possible by the late 2020s and likely by the early 2030s.

\begin{table}
\caption{Prospects for measuring the neutrino mass hierarchy in the coming years.  Sensitivity ranges are given to cover current uncertainties in other oscillation parameters.  Comments and concerns are given in the last column, with the most worrisome specific concerns placed in italics.}
\label{tab:summary}
\begin{center}
\begin{tabular}{@{}lccl@{}}
           & Hierarchy   & Approximate & \\
Experiment & sensitivity & timescale & Comments and concerns\\
\hline
\nova{}+T2K   & $1\text{--}3\sigma$     & 2020 & currently operating below designed beam power\\
DUNE      & $3\text{--}6\sigma$     & 2030 & {\em funding, timeline}\\
           & $5\text{--}10\sigma$    & 2035 & \\
PINGU      & $3\text{--}6\sigma$     & 2025 & funding; past systematics and resolution concerns largely addressed\\
ORCA       & $3\text{--}6\sigma$     & --   & insufficiently developed at present\\
Hyper-K    & $3\text{--}6\sigma$     & 2030 & {\em funding, timeline}\\
ICAL@INO   & $2\text{--}4\sigma$     & 2030 & timeline\\
JUNO       & $\mathord{\sim}4\sigma$ & 2027 & {\em detector performance not yet demonstrated}\\
RENO-50    & $\mathord{\sim}3\sigma$ & --   & insufficiently developed at present\\
Cosmology  & $0\text{--}4\sigma$     & 2027 & $0\sigma$ for most of allowed mass range; {\em requires minimal NH spectrum}\\
\hline
\end{tabular}
\end{center}
\end{table}

\end{document}